\documentclass[]{emulateapj}
\usepackage{apjfonts}

\slugcomment{}

\shorttitle{Mass function of low mass dark halos}
\shortauthors{Yahagi, Nagashima, \& Yoshii}

\begin{document}

\title{Mass Function of Low Mass Dark Halos}

\author{Hideki Yahagi\altaffilmark{1}}
\affil{Division of Theoretical Astrophysics,
	National Astronomical Observatory of Japan, Tokyo 181-8588, Japan}
\email{hideki.yahagi@nao.ac.jp}

\author{Masahiro Nagashima}
\affil{Department of Physics, University of Durham,
	South Road, Durham DH1 3LE, U. K.}
\email{masahiro.nagashima@durham.ac.uk}

\and

\author{Yuzuru Yoshii\altaffilmark{2}}
\affil{Institute of Astronomy, University of Tokyo, 2-21-1 Osawa,
Mitaka, Tokyo 181-0015, Japan}

\altaffiltext{1}{NAOJ Science Research Fellow}
\altaffiltext{2}{Research Center for the Early Universe,
	University of Tokyo, 7-3-1 Hongo, Bunkyo-ku, Tokyo 113-0033, Japan}

\begin{abstract}
The mass function of dark halos in a $\Lambda$-dominated cold dark
matter ($\Lambda$CDM) universe is investigated.  529 output files from five
runs of $N$-body simulations are analyzed using the friends-of-friends
cluster finding algorithm.  All the runs use $512^3$
particles in the box size of 35 $h^{-1}$Mpc to 140 $h^{-1}$Mpc.
Mass of particles for 35 $h^{-1}$ Mpc runs is 2.67 $\times 10^7
h^{-1}$ $M_{\odot}$.  Because of the high mass resolution of our
simulations, the multiplicity function in the low-mass range, where the
mass is well below the characteristic mass and $\nu = \delta_c /
\sigma \lesssim 1.0$, is evaluated in the present work, and
is well fitted by the functional form
proposed by Sheth \& Tormen (ST).  However, the maximum value of the
multiplicity function from our simulations at $\nu \sim 1$ is smaller,
and its low mass tail is shallower when
compared with the ST multiplicity function.
\end{abstract}


\section{Introduction}

The mass function is one of the most important statistical quantities
of dark
halos.  \citet{ps74} invented an ansatz to predict the mass function
of dark halos formed through hierarchical clustering, assuming that
each smoothed mass element with arbitrary smoothing length evolves
independently, in accordance with the spherical top-hat model.  The PS
formalism was extended so that the merging history of each halo is
traceable \citep{bcek91, bwr91, lc93}.  This extended PS formalism has
been used in a wide range of applications. Especially the so-called
semi-analytic galaxy models \citep{kwg93, sp99, cole00, ntgy01} use it
to successfully reproduce many properties of observed galaxies

The PS mass function and the mass function from $N$-body simulations
agree with each other only qualitatively \citep{efwd88, bv92}, and
modification to threshold linear overdensity leads to a better 
agreement \citep{er88, lc94, gb94}.  However, the PS mass function
gives the smaller number of high-mass halos \citep{jb94, gsphk98},
while giving the larger number of low-mass halos
when compared with the $N$-body mass functions \citep{gov99, slkd00}.
This tendency was confirmed over a mass range by connecting the
numerical mass functions from simulations with different box sizes
\citep{gsphk98, j01}.  Taking into account the spatial correlation of
density fluctuations \citep{yng96, ngsm01}, or incorporating the
ellipsoidal collapse model into the PS ansatz instead of the spherical
collapse model \citep{eps84, mnc95, ata97, ls98, smt01, st02},
an agreement was improved between the numerical mass function and the
analytic mass function, especially the ST mass function proposed by
\citet{st99}.  In addition, the mass function of
cluster progenitors has also been studied \citep{oh99}.

\defcitealias{j01}{J01}
\defcitealias{wht02}{W02}

On the other hand, the numerical mass function depends on the cluster
finding algorithm adopted \citep{lc93,gb94,gov99}.
\citet[hereafter J01]{j01} demonstrated that
the friends-of-friends (FoF) algorithm \citep{fof}
with a fixed linking length of 0.2 times the mean particle separation
results in the numerical mass function that shows the best
universality for various compositions of cosmological parameters
and box sizes.  Thus, it is better
to set the threshold density proportional to the {\it background} mass
density rather than to the {\it critical} density.  
\citet[hereafter W02]{wht02} supported the above argument and
demonstrated that an other definition of mass of halos, $M_{180b}$,
gives the universal numerical mass function. Here, $M_{180b}$ is the
mass within a sphere whose average density is 180 times the
{\it background} density.

However, different authors support different mass functions.  
For example, \citetalias{j01} proposed a new fitting formula to the
universal mass function, while
%%%%%%%%%%%%%%%%%%%%%%%%%%%%%%%%%%%%%%%%%%%%%%%%%%
%QUESTION 1 related sentence modification
\citetalias{wht02} supported the ST mass function
\citep[see also][]{reed03}.  Since
the mass function of high-mass halos has extensively been studied by
them, this discrepancy is possibly due to the lack of information on
the behavior of the numerical mass function for low-mass halos.  Thus,
in order to investigate the functional form of the universal mass
function, we performed five runs of $N$-body simulations with high
mass resolution.

In this paper, we briefly describe our simulation code and
cosmological and simulation parameters adopted in \S 2.  Results of
the simulations as well as the parameter values for the best-fit
mass function are given in \S 3, and discussions are given in \S4.

\section{Multiplicity function}
The multiplicity function is the differential distribution function of
the normalized fluctuation amplitude of dark halos for each mass
element. According to the definition by \citet{st99}, the
multiplicity function is defined as
\begin{eqnarray}
\nu f(\nu) &=& M^2 \frac{n(M, z)}{\bar \rho}
	\frac{d \log M}{d \log \nu},
\end{eqnarray}
where $n(M, z)$ is the number density of dark halos, 
$\nu = \delta_c / \sigma_M$ is the peak height of a halo,  $\delta_c$ is the
linear overdensity at the collapse epoch of halos given by the spherical
collapse model, and $\sigma_M$ is the standard deviation of the
density fluctuation field smoothed by
the top-hat window function.  Note that the time evolution of the
number density and the dependence of the number density on the
initial power spectrum are absorbed in $\nu$.  Since this
universality of the multiplicity function is guaranteed under the PS
ansatz, we compare our simulation results with analytic multiplicity
functions.  

Along with the PS ansatz, \citet{ps74} proposed the following
analytic multiplicity function:
\begin{eqnarray}
\mbox{PS:} \quad
\nu f(\nu) = \sqrt{\frac{2}{\pi}} \nu \exp(-\nu^2/2).
\end{eqnarray}
This PS multiplicity function has successfully predicted the
numerical multiplicity function in a qualitative manner.  However,
\citet{st99} proposed an alternative analytic multiplicity function
that could better reproduce the numerical multiplicity function:
\begin{eqnarray}
\mbox{ST:} \quad
\nu f(\nu)= A (1 + \nu'^{-2p}) \sqrt{\frac{2}{\pi}} \nu'
\exp(-\nu'^2/2),
\label{eq:ST}
\end{eqnarray}
where $\nu'^2 = a \nu^2$, and $A$ is a normalization factor defined so
that $\int_0^{\infty}f(\nu)d\nu=1$.  Because of this unity constraint,
$A$ is not an independent parameter, but is expressed in forms of $p$:
\begin{eqnarray}
A = \left[1 + 2^{-p}\pi^{-1/2}\Gamma(1/2 - p)\right]^{-1}.
\end{eqnarray}
\citet{st99} gave the best-fit parameter values of $a=0.707, p=0.3,$ and
$A=0.322$.  The PS multiplicity function is included in this ST
multiplicity function with $a=1, p=0$, and $A=1/2$.

The ST multiplicity function has a maximum at $\nu=\nu_{\mbox{max}}$
that satisfies the following equation: 
\begin{eqnarray}
(\nu'_{\mbox{max}}{}^{2p}+1)(\nu'_{\mbox{max}}{}^2-1)+2p=0,\label{eq:numax}
\end{eqnarray}
where $\nu'_{\mbox{max}}{}^2=a\nu_{\mbox{max}}{}^2$.  It is trivial to see
$\nu_{\mbox{max}}=1$ for the case of the PS multiplicity function.

\citetalias{j01} also proposed an analytic multiplicity function which
gives a fit to their numerical multiplicity function:
\begin{eqnarray}
\mbox{J01:} \quad
\nu f(\nu) = 0.315 \exp(-|0.61 + \log\nu - \log\delta_c|^{3.8}).
\end{eqnarray}
The valid range of this fitting formula is $-1.2 < \log\nu -
\log\delta_c < 1.05$.  Hereafter we assume that $\delta_{c}$ in the
above equation is constant and takes the value in the Einstein-de
Sitter universe, i.e. $\delta_c = (3/20)(12\pi)^{2/3} \sim 1.686$ in
order to compare the J01 multiplicity function with other functions.

\section{Simulations and Results}
We used the Adaptive Mesh Refinement $N$-body code developed by
\citet{yahagi}, which is a  vectorized and parallelized version of the
code described in \citet{yy01}.  All five runs of simulations we
performed adopt the
$\Lambda$CDM cosmological parameters of $\Omega_m=0.3$,
$\Omega_\lambda=0.7$, $h=0.7$, and $\sigma_8=1.0$, using
$512^3$ particles in common. The size of the finest mesh is 1/64 of
the base mesh, and the force dynamic range is
$2^{15} = 32768$.  Other simulation parameters, such as the box size
and the particle mass are given in Table \ref{tab:sim}.  Initial
conditions were
generated by the {\tt GRAFIC2} code provided by \citet{grafic} using
the power spectrum given by \citet{bbks}.  Five runs
produced 529 files, and each of them was analyzed by the FoF algorithm
with a constant linking length of 0.2 times the mean particle
separation.  Details of the simulations will be given in Yahagi et
al., in preparation.

\begin{figure}
\epsscale{.60}
\plotone{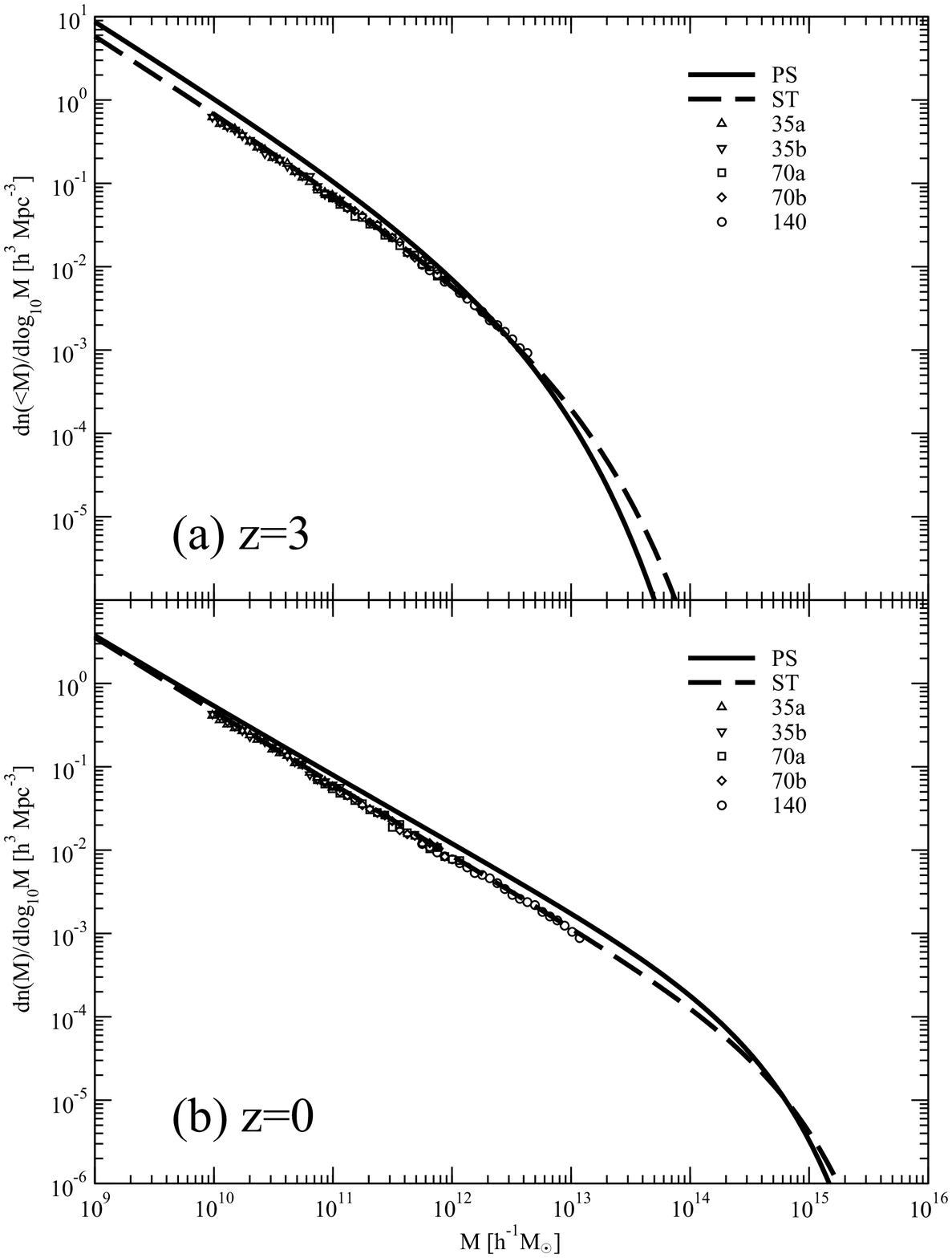}
\caption{
The mass functions of dark halos at ($a$) $z=3$ and ($b$) $z=0$.  The ST
mass function ({\it dashed line}) agrees with the numerical mass functions
({\it symbols}) fairly well at both redshifts, while the PS mass function
({\it solid line}) does not agree with them.
\label{fig:mf}}
\end{figure}

The mass functions of dark halos are shown in Figure \ref{fig:mf}.
The upper and lower panels show the mass functions at $z=3$ and $z=0$,
respectively.  The PS mass function is 
represented by solid lines, and the ST mass function  by dashed
lines.  At both redshifts, the numerical mass functions from our
simulations agree
with the ST mass function in a mass range of $10^{10} M_{\odot}
\lesssim M \lesssim 10^{13} M_{\odot}$.

\begin{figure}
\plotone{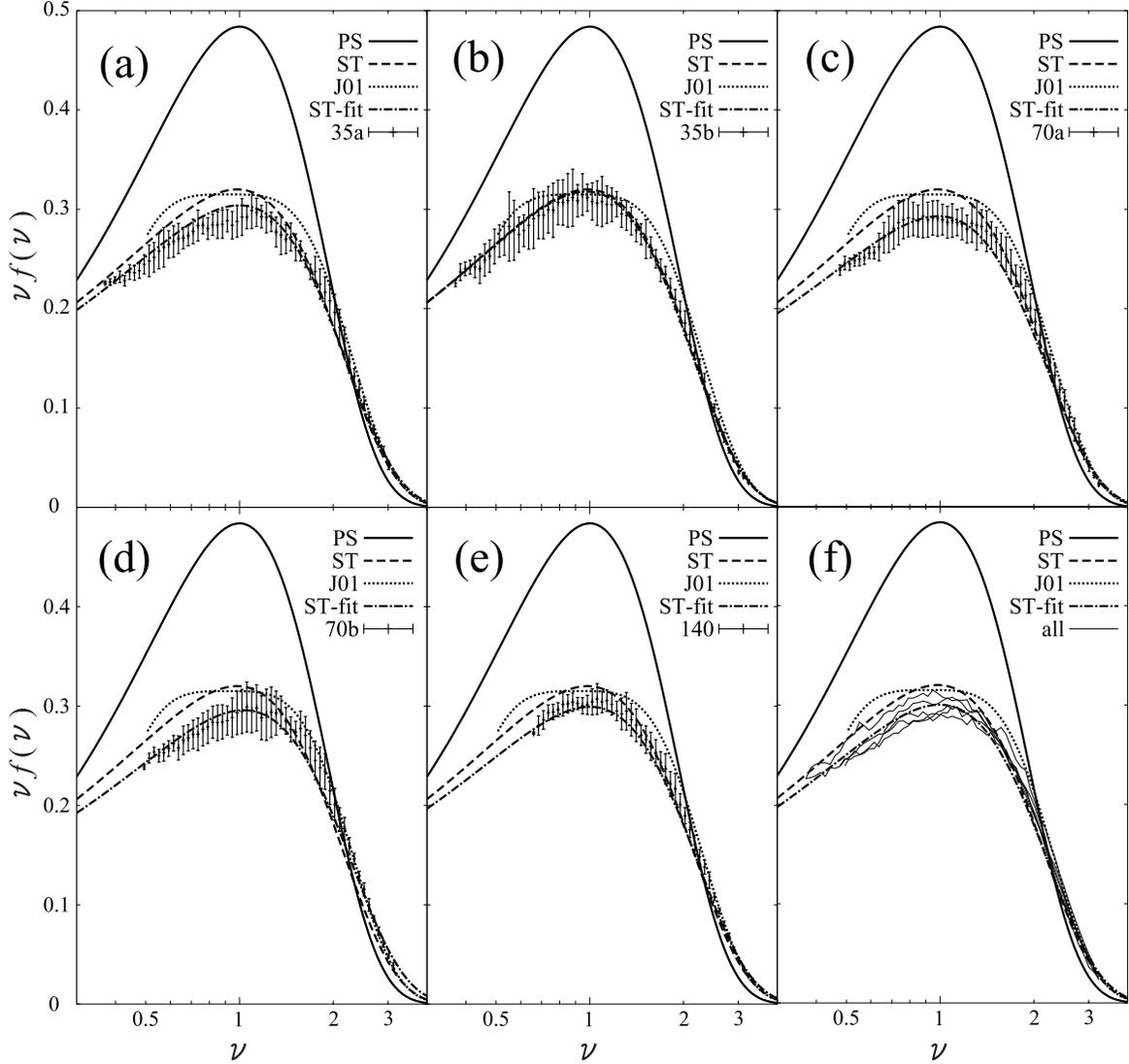}
\caption{
The numerical multiplicity functions from five runs of our simulations ({\it
crosses with error bars}) are shown in the panels ($a$-$e$),
except for the panel ($f$) which shows the results from all the runs
({\it thin lines}).
%%%%%%%%%%%%%%%%%%%%%%%%%%%%%%%%%%%%%%%%%%%%%%%%%%
%QUESTION 2 related sentence insertion
The numerical multiplicity functions are derived by compiling all the
data from the initial redshift to the present $z=0$. Crosses and 
error bars represent their average and rms, respectively.
%%%%%%%%%%%%%%%%%%%%%%%%%%%%%%%%%%%%%%%%%%%%%%%%%%
Also shown in each panel are the PS multiplicity
function ({\it solid line}), the ST multiplicity function ({\it dashed
line}), the J01 multiplicity function ({\it dotted line}), and the best-fit
function using the ST functional form ({\it dot-dashed
line}) for which adopted parameters are given in Table
\ref{tab:fit}.  The J01 multiplicity function, originally given as
a function of $\sigma$ \citep{j01}, is expressed here in terms of $\nu
= \delta_c / \sigma$, assuming that $\delta_c$ is constant although it
varies slightly in the $\Lambda$CDM universe. In the high-mass range
($\nu>$1), the numerical multiplicity functions reside
between the ST and J01 functions, and its maximum value at
$\nu \sim 1$ is below those of the ST and J01 functions.
\label{fig:mltpl}}
\end{figure}

The numerical multiplicity functions are shown
by crosses in five panels of Figure \ref{fig:mltpl}.
%%%%%%%%%%%%%%%%%%%%%%%%%%%%%%%%%%%%%%%%%%%%%%%%%%
%Question 2 related sentence insertion
All the data from the initial redshift to the present $z=0$ is
compiled to draw the average curves ({\it crosses}) with 
error bars indicating the epoch to epoch variation.
%%%%%%%%%%%%%%%%%%%%%%%%%%%%%%%%%%%%%%%%%%%%%%%%%%
In the panel (f), all the numerical multiplicity
functions are shown by thin lines.  Dark halos that consist of less
than 600 particles are not used in calculating the multiplicity
function, and 1/64 dex-sized bins containing less than 100 halos are
excluded to avoid the contamination of the rare objects.  Three analytic
multiplicity functions described in the previous section are also
shown in this figure, that is PS ({\it solid lines}), ST ({\it dashed
lines}), and J01 ({\it dotted lines}).
The best-fit functions based on the ST functional form
are also shown by dot-dashed lines.  The best-fit parameters are given
in Table \ref{tab:fit} and 
are very close to those of
\citetalias{wht02}.  Since the data are available only in the region at
$\nu \lesssim 3$, these functions could be erroneous at $\nu \gtrsim 3$.

\begin{figure}
\plotone{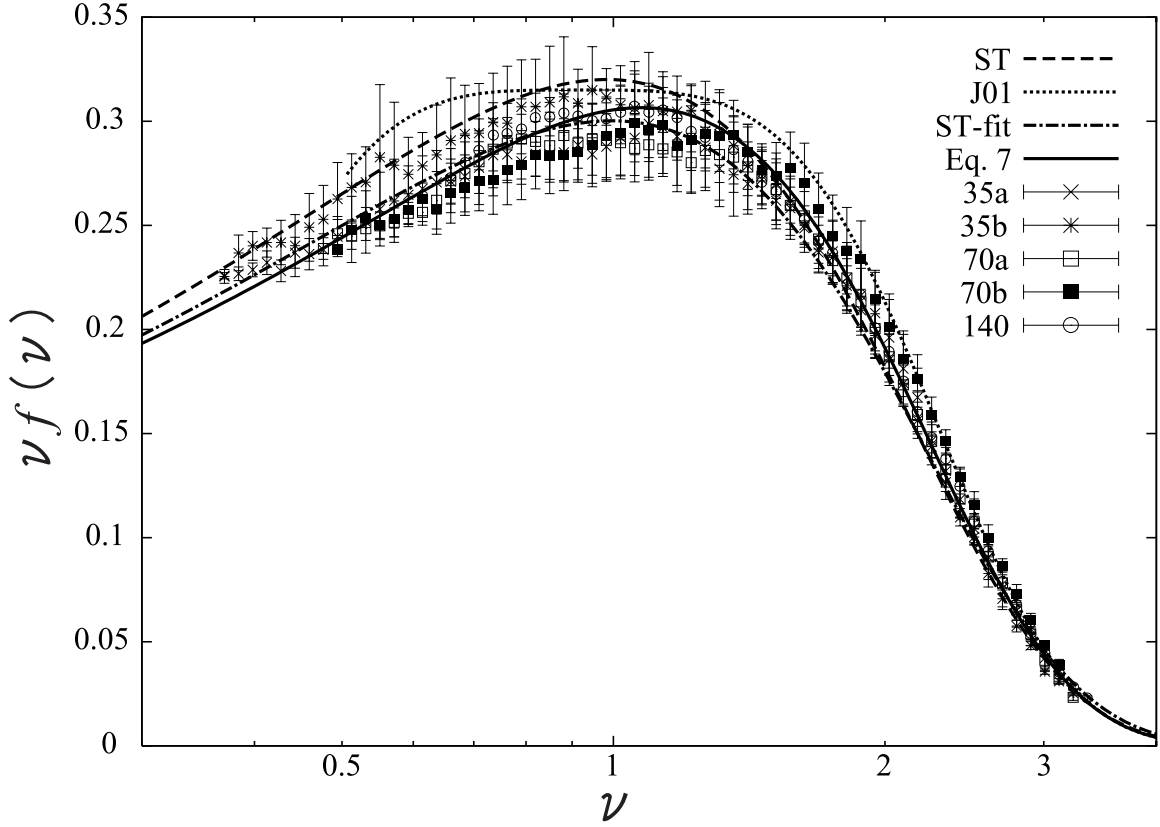}
\caption{
The best-fit multiplicity function using equation \ref{eq:4fit} 
({\it solid line}).  This function well represents the numerical multiplicity
functions indicated by symbols with error bars.
Also shown for comparison are the ST ({\it dashed line}), J01 ({\it
dotted line}), and best-fit ST ({\it dot-dashed line}) functions.
For $1.5 \lesssim \nu \lesssim 3$, all the numerical
functions reside between the ST and the J01
functions.  Even the best-fit ST function cannot represent
the multiplicity functions well in this region.
\label{fig:4fit}}
\end{figure}

In most cases, the numerical multiplicity functions and the best-fit
functions to them are consistent with the ST and J01 multiplicity
functions at $\nu \gtrsim 3$.  However, each of the numerical multiplicity
functions reside between the ST and J01 functions at $1.5
\lesssim \nu \lesssim 3$, and is below the ST
function at $\nu \lesssim 1$ except for the 35b run.  The
numerical multiplicity functions have an apparent peak at $\nu
\sim 1$, instead of a plateau as seen in the J01 function.  We here
proposed the following function to fit to the numerical multiplicity
function:
\begin{eqnarray}
%\nu f(\nu) = A [1+(B \nu / \sqrt{2})^C] \nu^D \exp[-(B \nu)^2/2]
\nu f(\nu) = A [1+(B \nu / \sqrt{2})^C] \nu^D \exp[-(B \nu/\sqrt{2})^2],
\label{eq:4fit}
\end{eqnarray}
%%%%%%%%%%%%%%%%%%%%%%%%%%%%%%%%%%%%%%%%%%%%%%%%%%
%QUESTION 5 related sentence modification
%Here, $A$ is not a free parameter, but a normalization factor to
%satisfy $\int_0^{\infty}f(\nu)d\nu=1$.  Hence, 
where, $A$ is a normalization factor to satisfy the unity constraint,
$\int_0^{\infty}f(\nu)d\nu=1$, therefore
\begin{eqnarray}
A=2 (B/\sqrt{2})^D\{\Gamma[D/2] + \Gamma[(C+D)/2]\}^{-1}.
\end{eqnarray}
The best-fit parameters are given
as $B$=0.893, $C$=1.39, and $D$=0.408, and from these parameters, $A$
is constrained so that $A=0.298$.  This best-fit function from
equation \ref{eq:4fit} is shown in
Figure \ref{fig:4fit} and is only valid at $0.3 \leq \nu \leq 3$.

%%%%%%%%%%%%%%%%%%%%%%%%%%%%%%%%%%%%%%%%%%%%%%%%%%
%QUESTION 5 related paragraph insertion
Since the unity constraint is only an assumption,
we can relax this constraint and treat $A$ as a free parameter.
Actually, almost
a half of the particles are not bound to any halos as shown in Table
\ref{tab:unbnd}.  On the other hand, the best-fit parameters without
the unity constraint using equation \ref{fig:4fit} are
$A=0.320$, $B=0.664$, $C=1.99$, and $D=0.36$, for which
$\int_0^{\infty}f(\nu)d\nu=1.2713$.  Since the fraction of bound
particle should not exceed unity, we assign the unity constraint to
all the best-fit functions below.
%%%%%%%%%%%%%%%%%%%%%%%%%%%%%%%%%%%%%%%%%%%%%%%%%%

We also investigate the peak position of the multiplicity functions.
Since deriving $\nu_{\mbox{max}}$ directly from the numerical
multiplicity function is difficult due to the numerical fluctuation
around the maximum value, we derived $\nu_{\mbox{max}}$ from the
best fit ST function by solving equation \ref{eq:numax}.  These
$\nu_{\mbox{max}}$ are given in Table \ref{tab:fit}, and are
very close to unity.  For reference, $\nu_{\mbox{max}}=0.916$ for
the J01 multiplicity function, and $\nu_{\mbox{max}}=0.881$ for the
multiplicity function proposed by \citet{ls98}.

\begin{figure}
\plotone{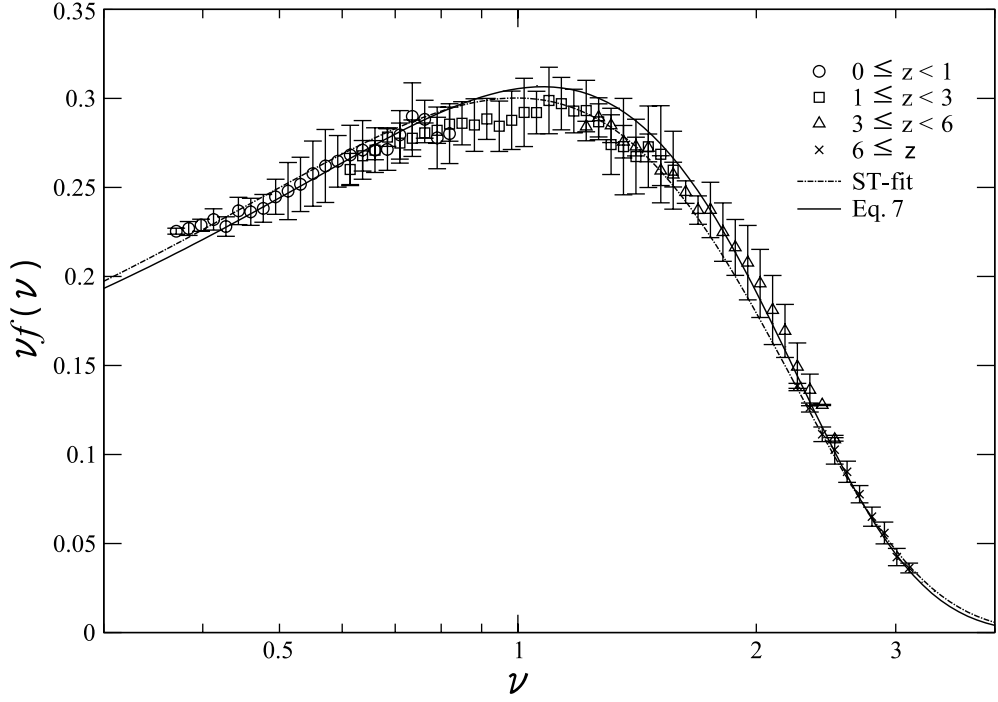}
\caption{
Time dependence of the multiplicity function from the 35a run;
$0 \leq z < 1$ ({\it circles}),
$1 \leq z < 3$ ({\it squares}),
$3 \leq z < 6$ ({\it triangles}), and
$z \geq 6$     ({\it crosses}).
Also shown are the best fit function for all the runs using the ST
functional form
({\it dot-dashed line}) and  equation \ref{eq:4fit} ({\it solid line}).
At high redshifts, high-$\nu$ halos in the exponential
part of the ST and equation \ref{eq:4fit} are probed.  As redshift
decreases, the probe window moves to the lower-$\nu$ region.
\label{fig:timedep}}
\end{figure}

%%%%%%%%%%%%%%%%%%%%%%%%%%%%%%%%%%%%%%%%%%%%%%%%%%
%QUESTION 2 related paragraph insertion
We also checked the time dependence of the multiplicity function.
Figure \ref{fig:timedep} shows the multiplicity
function from the 35a run, for four redshift ranges of 
$0 \leq z < 1$ ({\it circles}),
$1 \leq z < 3$ ({\it squares}),
$3 \leq z < 6$ ({\it triangles}), and
$z \geq 6$,    ({\it crosses}).
At high redshifts, high-$\nu$ halos in the exponential part of the
best-fit ST
function and equation \ref{eq:4fit} are probed.  As redshift decreases,
the probe window moves to the lower-$\nu$ region.
%%%%%%%%%%%%%%%%%%%%%%%%%%%%%%%%%%%%%%%%%%%%%%%%%%

Since there are no modes of density fluctuations whose wave length
is larger than the box size on which the periodic boundary is placed. The
numerical multiplicity function is conditional such that
$f(\nu|\delta_{box}=0)$,
where $\delta_{box}$ is the density contrast smoothed over a
mass scale comparable to the box size.
Since our box size is smaller than that of other
groups, we have estimated this box-size effect
%%%%%%%%%%%%%%%%%%%%%%%%%%%%%%%%%%%%%%%%%%%%%%%%%%
%QUESTION 3 related equation insertion
% using the PS
%conditional multiplicity function.  This effect is not so large where
comparing the unconditional PS multiplicity function using the sharp
$k$-space filter with the conditional one \citep{bcek91, bwr91, lc93},
\begin{eqnarray}
\nu f_c=\sqrt\frac{2}{\pi}
\frac{\nu-\epsilon\nu_{box}}{(1-\epsilon^2)^{3/2}}
\exp{\left[-\frac{(\nu-\epsilon\nu_{box})^2}{2(1-\epsilon^2)}\right]},
\label{eq:cndPS}
\end{eqnarray}
where $\epsilon=\sigma_{box}/\sigma$. Setting $\nu_{box}=0$, 
this effect is found to be not so large at 
$\nu \lesssim 3$ for the box size of 35$h^{-1}$Mpc, and 
it makes the universality of the multiplicity function even worse.

\section{Discussion}
We have performed five runs of $N$-body simulations with high mass
resolution in
order to study the behavior of the numerical multiplicity function in
the low-mass range, or the low-$\nu$ region.  Throughout the
peak range of,
$0.3 \leq \nu \leq 3$, the ST functional form provides a
good fit to them with parameter values of $a=0.664, p=0.321$, and
$A=0.301$.  These values are very close to those of
\citetalias{wht02}.  Our numerical multiplicity functions have
a peak at $\nu \sim 1$ as in the ST function, instead of a plateau
in the J01 function.

%However, there are some discrepancies between our numerical
%multiplicity function and model functions.  First, in the low $\nu$
However, some detailed discrepancies are seen between
our numerical multiplicity functions and the ST and J01 analytic
functions.  First, in the low-$\nu$ region of $\nu \lesssim 1$, our
numerical multiplicity functions systematically fall below the ST and
the J01 functions, while they are consistent with that of
%%%%%%%%%%%%%%%%%%%%%%%%%%%%%%%%%%%%%%%%%%%%%%%%%%
%QUESTION 6 related sentence removal.
%\citetalias{wht02}.  This discrepancy is due to the the number of
%data used to fit.  On the other hand, in the high $\nu$ region, where
\citetalias{wht02}.
On the other hand, in the high-$\nu$ region, where
%%%%%%%%%%%%%%%%%%%%%%%%%%%%%%%%%%%%%%%%%%%%%%%%%%
$\nu$ is significantly larger
than unity, our numerical multiplicity functions take values between the ST
and J01 functions.  Although these differences are
within 1 $\sigma$ error bars, they are possibly due
to the different box sizes adopted. \citet{st99} used the data from the
GIF simulations \citep{gif99} with the box size of 144$h^{-1}$Mpc or less,
and \citetalias{wht02} mainly used the data from the box size of
200$h^{-1}$Mpc.  On the other hand, \citetalias{j01} used the
3000 $h^{-1}$Mpc simulation at maximum.  We have taken into account
the box-size effect using the conditional PS multiplicity function
(equation \ref{eq:cndPS})
but this effect is found to be too small to resolve this problem.
Introducing the estimation using the conditional multiplicity
function based on the unconditional multiplicity function which fits
the numerical multiplicity function well, such as the ST multiplicity
function, would resolve this problem. However, such an improved
estimation of the box-size effect might be weaker than
the estimation based on the PS function, because 
the unconditional ST function at $\nu\sim1$ has a broad
peak that is below that of the PS function.  The fact that our
numerical multiplicity functions keep the universality supports this
line of argument.

%Thus, there are two discrepancies remained.  First one is the
%discrepancy between the numerical multiplicity function and the model
%numerical multiplicity function.  We require a new model function
%which fit numerical multiplicity function better keeping some required
%conditions such as, $\int_0^{\infty}f(\nu)\mbox{d}\nu=1$.
Thus, there are two discrepancies remained.  One is the
discrepancy between the numerical and analytical multiplicity functions.
Although our newly proposed functional form
(equation \ref{eq:4fit}) provides a better fit when compared
with the ST functional form (equation \ref{eq:ST}),
  we need an analytic function based on a
theoretical background which fits the numerical multiplicity function
even better.
The other one
is the discrepancy in the numerical multiplicity functions from various
simulation runs.  There are three strategies to resolve this
discrepancy.
The first is to run simulations having still higher mass dynamic
range free from the box size effect.  The second is to increase the
number of
realizations as \citetalias{wht02} did, because there
is a scatter from the runs using the same box size.  The third
is to run simulations whose box size is {\it smaller} than
that of the present work, although it might sound contradictorily.  
From simulations with smaller box size, we will
obtain the information
on the conditional multiplicity function which coincides with the
unconditional multiplicity function at $\nu \ll 1$.  Comparing the
unconditional multiplicity function from simulations with a large box size
and the conditional multiplicity function from those of a small box size
will offer not only the clues to resolve the above
mentioned discrepancies, but also insights into the mechanism how the
PS ansatz works to reproduce the numerical multiplicity function.

\acknowledgments
Simulations described in this paper were carried out using
Fujitsu-made vector parallel processors VPP5000 installed at the
Astronomical Data Analysis Center, National Astronomical Observatory
of Japan (ADAC/NAOJ), under the ADAC/NAOJ large scale simulation
projects (group-ID: myy26a, yhy35b).  HY would like to thank Joseph
Silk and Masahiro Takada for their useful comments.  MN acknowledges
support from a PPARC rolling grant for extragalactic astronomy and
cosmology.
This work has been supported partly by the Center of Excellence
Research (07CE2002) of the Ministry of Education, Science, Culture,
and Sports of Japan.

\clearpage

\clearpage 

\clearpage 

\clearpage 

\clearpage

%\begin{deluxetable}{lccccc}
%\tablecolumns{6}
%\tablewidth{0pc}
%\tablecaption{Characteristic of Simulations in the Present Work\label{tab:param}}
%\tablehead{
%\colhead{} & \multicolumn{3}{c}{Simulation Parameters} & \colhead{} &
%\colhead{Unbound particles at $z=0$}\\
%\cline{2-4} \cline{6-6}\\
%\colhead{Model} &
%\colhead{$L$ [$h^{-1}$Mpc]} &
%\colhead{$m_{ptcl}$ [$h^{-1} M_\odot$]} &
%\colhead{$z_{\mbox{start}}$} &
%\colhead{} &
%\colhead{$x_{\mbox{ub}}$\tablenotemark{a}}
%}
%\startdata
%35a\dotfill &35 & $2.67 \times 10^7$ & 50 & \phn & 0.472\\
%35b\dotfill &35 & $2.67 \times 10^7$ & 50 & \phn & 0.472\\
%70a\dotfill &70 & $2.13 \times 10^8$ & 41 & \phn & 0.530\\
%70b\dotfill &70 & $2.13 \times 10^8$ & 41 & \phn & 0.528\\
%140\dotfill &140& $1.70 \times 10^9$ & 33 & \phn & 0.601\\
%\enddata
%\tablecomments{All the simulations adopt the cosmological parameters of
%$\Omega_m=0.3$, $\Omega_\lambda=0.7$, $h=0.7$, and $\sigma_8=1.0$.
%The number of particles is $512^3$ and the force dynamic range of
%simulations is $2^{15}=32768$ in common. $L$ is the box size,
%$m_{ptcl}$ is the particle mass, and $z_{\mbox{start}}$ is the initial
%redshift.}
%\tablenotetext{a}{The fraction of unbound particles in number.}
%\end{deluxetable}

\begin{deluxetable}{lccc}
\tabletypesize{\scriptsize}
\tablecaption{Simulation Parameters\label{tab:sim}}
\tablewidth{0pt}
\tablehead{
\colhead{Model} &
\colhead{$L$ [$h^{-1}$Mpc]}&
\colhead{$m_{ptcl}$ [$h^{-1}$M$_\odot$]}&
\colhead{$z_{\mbox{start}}$}
}
\startdata
35a\dotfill &35 & $2.67 \times 10^7$ & 50\\
35b\dotfill &35 & $2.67 \times 10^7$ & 50\\
70a\dotfill &70 & $2.13 \times 10^8$ & 41\\
70b\dotfill &70 & $2.13 \times 10^8$ & 41\\
140\dotfill &140& $1.70 \times 10^9$ & 33
\enddata

\tablecomments{All the simulations adopt the cosmological parameters of
$\Omega_m=0.3$, $\Omega_\lambda=0.7$, $h=0.7$, and $\sigma_8=1.0$.
The number of particles is $512^3$ and the force dynamic range of
simulations is $2^{15}=32768$ in common. $L$ is the box size,
$m_{ptcl}$ is the particle mass, and $z_{\mbox{start}}$ is the initial
redshift.}
\end{deluxetable}

\clearpage

\begin{deluxetable}{cllll}
\tabletypesize{\scriptsize}
\tablecaption{Best-Fit Parameters\label{tab:fit}}
\tablewidth{0pt}
\tablehead{
\colhead{Model}& \colhead{$a$}   & \colhead{$p$} & \colhead{$A$} &
\colhead{$\nu_{\mbox{max}}$\tablenotemark{a}}
}
\startdata
35a\dotfill & 0.665 & 0.317 & 0.305 & 0.998\\
35b\dotfill & 0.715 & 0.303 & 0.320 & 0.975\\
70a\dotfill & 0.666 & 0.327 & 0.293 & 0.988\\
70b\dotfill & 0.614 & 0.325 & 0.296 & 1.031\\
140\dotfill & 0.658 & 0.321 & 0.300 & 0.999\\
\\
all\dotfill & 0.664 & 0.321 & 0.301 & 0.996\\
\\
PS\dotfill & 1     & 0     & 0.5   & 1\\
ST\dotfill & 0.707 & 0.3   & 0.322 & 0.983\\
\citetalias{wht02}\dotfill & 0.64 & 0.34 & --- & 0.995
\enddata

\tablecomments{The ST functional form in equation \ref{eq:ST} is used
to fit to the numerical multiplicity function.}
\tablenotetext{a}{The value of $\nu$ at which the best-fit ST function
attains a maximum.}

\end{deluxetable}

\clearpage

%%%%%%%%%%%%%%%%%%%%%%%%%%%%%%%%%%%%%%%%%%%%%%%%%%
% QUESTION 5 related table addition
\begin{deluxetable}{cl}
\tabletypesize{\scriptsize}
\tablecaption{Fraction of Unbound Particles at $z$=0\label{tab:unbnd}}
\tablewidth{0pt}
\tablehead{
\colhead{Model}& \colhead{$x_{\mbox{ub}}$}\tablenotemark{a}}
\startdata
35a\dotfill & 0.472\\
35b\dotfill & 0.472\\
70a\dotfill & 0.530\\
70b\dotfill & 0.528\\
140\dotfill & 0.601
\enddata
\tablenotetext{a}{Fraction of unbound particles.}
\end{deluxetable}
%%%%%%%%%%%%%%%%%%%%%%%%%%%%%%%%%%%%%%%%%%%%%%%%%%


\begin{thebibliography}{}
\bibitem[Audit, Teyssier, \& Alimi(1997)]{ata97} Audit, E., Teyssier, R.
\& Alimi, J.-M. 1997, \aap, 325, 439
\bibitem[Bardeen et al.(1986)]{bbks} Bardeen, J. M., Bond, J. R.,
Kaiser, N., \& Szalay, A. S. 1986, \apj, 304, 15
\bibitem[Bertschinger(2001)]{grafic} Bertschinger, E. 2001, \apjs,
137, 1 
\bibitem[Bond et al.(1991)]{bcek91} Bond, J. R., Cole, S., Efstathiou,
G., \& Kaiser, N. 1991, \apj, 379, 440
\bibitem[Bower(1991)]{bwr91} Bower, R. G. 1991 \mnras, 248, 332
\bibitem[Brainerd \& Villumsen(1992)]{bv92} Brainerd, T. G., \&
Villumsen, J. V. 1992, \apj, 394, 409 
\bibitem[Cole et al.(2000)]{cole00} Cole, S., Lacey, C. G., Baugh,
C. M., \& Frenk, C. S. 2000, \mnras, 319, 168
\bibitem[Davis et al.(1985)]{fof} Davis, M., Efstathiou, G., Frenk,
C. S., \& White, S. D. M. 1985, \apj, 292, 371 
\bibitem[Efstathiou et al.(1988)]{efwd88} Efstathiou, G.,  Frenk, C. S.,
White, S. D. M., \& Davis, M. 1988, \mnras, 235, 715
\bibitem[Efstathiou \& Rees(1988)]{er88} Efstathiou, G., \& Rees,
M. J. 1988, \mnras, 230, 5p
\bibitem[Epstein(1984)]{eps84} Epstein, R. I. 1984, \apj, 281, 545
\bibitem[Gelb \& Bertschinger(1994)]{gb94} Gelb, J. M., \&
Bertschinger, E. 1994, \apj, 436, 467
\bibitem[Governato et al.(1999)]{gov99} Governato, F., Babul, A.,
Quinn, T., Tozzi, P., Baugh, C. M., Katz, N., \& Lake, G. 1999,
\mnras, 307, 949
\bibitem[Gross et al.(1998)]{gsphk98} Gross, M. A. K., Somerville,
R. S., Primack J. R., Holtzman, J., \& Klypin, A. 1998, \mnras, 301, 81
\bibitem[Jain \& Bertschinger(1994)]{jb94} Jain, B., \& Bertschinger,
E. 1994, \apj, 431, 495
\bibitem[Jenkins et al.(2001)]{j01} Jenkins, A., Frenk, C. S.,
White, S. D. M., Colberg, J. M., Cole, S., Evrard, A. E., Couchman,
H. M. P., \& Yoshida, N. 2001, \mnras, 321, 372
\bibitem[Kauffman et al.(1999)]{gif99} Kauffmann, G., Colberg, J. M.,
Diaferio, A., \& White, S. D. M. 1999, \mnras, 303, 188
\bibitem[Kauffmann, White, \& Guiderdoni(1993)]{kwg93} Kauffmann, G.,
White, S. D. M., \& Guiderdoni, B. 1993, \mnras, 264, 201  
\bibitem[Lacey \& Cole(1993)]{lc93} Lacey, C., \& Cole, S. 1993,
\mnras, 262, 627
\bibitem[Lacey \& Cole(1994)]{lc94} Lacey, C., \& Cole, S. 1994,
\mnras, 271, 676
\bibitem[Lee \& Shandarin(1998)]{ls98} Lee, J., \& Shandarin,
S. F. 1998, \apj, 500, 14
\bibitem[Monaco(1995)]{mnc95} Monaco, P. 1995, \apj, 447, 23
\bibitem[Nagashima(2001)]{ngsm01} Nagashima, M. 2001, \apj, 562, 7 
\bibitem[Nagashima et al.(2001)]{ntgy01} Nagashima, M., Totani, T.,
Gouda, N., \& Yoshii, Y. 2001, \apj, 557, 505
\bibitem[Okamoto \& Habe(1999)]{oh99} Okamoto, T., \& Habe, A. 1999,
\apj, 516, 591
\bibitem[Press \& Schechter(1974)]{ps74} Press, W. H., \& Schechter,
P. 1974, \apj, 187, 425
\bibitem[Reed et al.(2003)] {reed03} Reed, D., Gardner, J., Quinn, T.,
Stadel, J., Fardal, M., and Lake, G. 2003, preprint (astro-ph/0301270)
\bibitem[Sheth, Mo, \& Tormen(2001)]{smt01} Sheth, R. K., Mo, H. J.,
\& Tormen, G. 2001, \mnras, 323, 1
\bibitem[Sheth \& Tormen(1999)]{st99} Sheth, R. K., \& Tormen,
G. 1999, \mnras, 308, 119
\bibitem[Sheth \& Tormen(2002)]{st02} Sheth, R. K., \& Tormen,
G. 2002, \mnras, 329, 61
\bibitem[Somerville et al.(2000)]{slkd00} Somerville, R. S., Lemson,
G., Kolatt, T. S., \& Dekel, A. 2000, \mnras, 316, 479
\bibitem[Somerville \& Primack(1999)]{sp99} Somerville, R. S., \&
Primack J. R. 1999, \mnras, 310, 1087
\bibitem[White(2002)]{wht02} White, M. 2002, \apjs, 143, 241
\bibitem[Yahagi(2002)]{yahagi} Yahagi, H. 2002, Doctoral Thesis,
University of Tokyo
\bibitem[Yahagi \& Yoshii(2001)]{yy01} Yahagi, H., \& Yoshii, Y. 2001,
\apj, 558, 463
\bibitem[Yano, Nagashima, \& Gouda(1996)]{yng96} Yano, T., Nagashima,
M., \& Gouda, N. 1996, \apj, 466, 1

\end{thebibliography}
\end{document}